\documentclass[preprint]{emulateapj}
\def\la{\lambda}

\def\D{\Delta}

\usepackage[dvips]{color}

\usepackage{aas_macros} %added later
\usepackage{here}
\usepackage{amsmath,amssymb}
\slugcomment{}
\shortauthors{Hirano et al.}
\shorttitle{
Planet-Planet Eclipse and the Rossiter-McLaughlin Effect for KOI-94}
\begin{document}
%%%%%%%%%%%%%%%%%%%%%%%%%%%%%%%%%%%%%%%%%%%%%%%%%%%%%%%%%%%%%%%%%%%%%%
%%%%%%%%%%%%%%%%%%%%%%%%%%%%%%%%%%%%%%%%%%%%%%%%%%%%%%%%%%%%%%%%%%%%%%
%\title{Spin-orbit Alignment in the Multiple Transiting System KOI-94}
%\title{First Analysis of the Rossiter-McLaughlin Effect and 
%Planet-Planet Crossing Event in a Multiple Transiting System}
%\title{Discovery of the Planet-Planet Eclipse of a Multiple Transiting
%System KOI-94: Joint Analysis of the Rossiter-McLaughlin Effect and the
%Kepler Photometry}
\title{Planet-Planet Eclipse and the Rossiter-McLaughlin Effect 
 of a Multiple Transiting System: Joint Analysis of the Subaru 
 Spectroscopy and the Kepler Photometry}
%%%%%%%%%%%%%%%%%%%%%%%%%%%%%%%%%%%%%%%%%%%%%%%%%%%%%%%%%%%%%%%%%%%%%%
%%%%%%%%%%%%%%%%%%%%%%%%%%%%%%%%%%%%%%%%%%%%%%%%%%%%%%%%%%%%%%%%%%%%%%
\author{
Teruyuki \textsc{Hirano},\altaffilmark{1}
Norio \textsc{Narita},\altaffilmark{2}
Bun'ei \textsc{Sato},\altaffilmark{3}
Yasuhiro H.\ \textsc{Takahashi},\altaffilmark{2,4}
Kento \textsc{Masuda},\altaffilmark{1}
%Joshua N.\ \textsc{Winn},\altaffilmark{2}
Yoichi \textsc{Takeda},\altaffilmark{2}
Wako \textsc{Aoki},\altaffilmark{2}
Motohide \textsc{Tamura},\altaffilmark{2}
%Atsushi \textsc{Taruya}\altaffilmark{1}
and 
Yasushi \textsc{Suto}\altaffilmark{1,5}
} 
%%%%%%%%%%%%%%%%%%%%%%%%%%%%%%%%%%%%%%%%%%%%%%%%%%%%%%%%%%%%%%%%%%%%%%
\altaffiltext{1}{
Department of Physics, The University of Tokyo, Tokyo, 113-0033, Japan
}
%\altaffiltext{2}{
%Department of Physics, and Kavli Institute for Astrophysics and Space Research,\\
%Massachusetts Institute of Technology, Cambridge, MA 02139, USA
%}
\altaffiltext{2}{
National Astronomical Observatory of Japan, 2-21-1 Osawa,
Mitaka, Tokyo, 181-8588, Japan
}
\altaffiltext{3}{
Department of Earth and Planetary Sciences, Tokyo Institute of Technology,
2-12-1 Ookayama, Meguro-ku, Tokyo 152-8551, Japan
}
\altaffiltext{4}{
Department of Astronomy, The University of Tokyo, Tokyo, 113-0033, Japan
}
\altaffiltext{5}{Department of Astrophysical Sciences, Princeton
University, Princeton, NJ 08544}

\email{hirano@utap.phys.s.u-tokyo.ac.jp}
%%%%%%%%%%%%%%%%%%%%%%%%%%%%%%%%%%%%%%%%%%%%%%%%%%%%%%%%%%%%%%%%%%%%%%
\begin{abstract}
We report a joint analysis of the Rossiter-McLaughlin (RM) effect with
Subaru and the Kepler photometry for Kepler Object of Interest (KOI) 94
system. The system comprises four transiting planet candidates
with orbital periods of 22.3 (KOI-94.01), 10.4 (KOI-94.02),
54.3 (KOI-94.03), and 3.7 (KOI-94.04) days from the Kepler photometry.  
We performed the radial velocity (RV) measurement of the system with the
Subaru 8.2 m telescope on August 10, 2012 (UT), covering a complete transit
of KOI-94.01 for $\sim 6.7$ hours.  The resulting RV variation due to
the RM effect spectroscopically confirms that KOI-94.01 is indeed the
transiting planet, and implies that its orbital axis is well aligned
with the stellar spin axis; the projected spin-orbit angle $\lambda$ is
estimated as $-6_{-11}^{+13}$ deg.
This is the first measurement of the RM effect for a multiple transiting system. 
Remarkably, the archived Kepler lightcurve around BJD=2455211.5 (date in UT January 14/15, 2010) 
indicates a ``double transit'' event of KOI-94.01 and KOI-94.03, in
which the two planets transit the stellar disk simultaneously. Moreover,
the two planets partially overlap each other, and exhibit a
``planet-planet eclipse'' around the transit center. This provides a
rare opportunity to put tight constraints on the configuration of the two transiting planets
by joint analysis with our Subaru RM measurement.  Indeed, we find that
the projected mutual inclination of KOI-94.01 and KOI-94.03 is estimated
to be $\delta = -1.15^\circ \pm 0.55^\circ$.
Implications for the migration model of multiple planet systems are
also discussed.
\end{abstract}
%%%%%%%%%%%%%%%%%%%%%%%%%%%%%%%%%%%%%%%%%%%%%%%%%%%%%%%%%%%%%%%%%%%%%%
\keywords{planets and satellites: individual (KOI-94) -- 
planets and satellites: formation -- stars: rotation -- 
techniques: radial velocities -- 
techniques: spectroscopic}
%%%%%%%%%%%%%%%%%%%%%%%%%%%%%%%%%%%%%%%%%%%%%%%%%%%%%%%%%%%%%%%%%%%%%%
\section{Introduction \label{s:sec1}}\label{s:sec1}

The proximity of giant planets to their host stars implies that such
planets have gone through so-called planetary migrations.  The angle
between the stellar spin axis and planetary orbital axis is an important
probe to reveal the origin of planetary migrations 
\citep{2000A&A...359L..13Q, 2005ApJ...631.1215W}. 
%insight into the dynamical evolution of planetary systems. 
While migrations at earlier stages due to interactions between planets
and proto-planetary disk are supposed to result in smaller values for
the spin-orbit angles \citep{2010exop.book..347L}, migrations caused by
planet-planet scatterings or the Kozai cycle and subsequent tidal
interactions with the host star could produce large spin-orbit angles
\citep{2002Icar..156..570M, 2007ApJ...669.1298F, 2008ApJ...686..580C, 2011ApJ...742...72N}.

Observations of the Rossiter-McLaughlin (RM) effect have been the major
channel to measure the sky-projected spin-orbit angles $\la$
\citep[e.g., ][]{2005ApJ...622.1118O, 2007PASJ...59..763N,
2012ApJ...757...18A}.  With previous dedicated campaigns, last few years
have
witnessed an unexpectedly large fraction of spin-orbit misaligned
systems, and even retrograde orbits with respect to the projected
stellar equator have been reported \citep{2009PASJ...61L..35N,
2009ApJ...703L..99W, 2010A&A...524A..25T}, indicating that dynamical
processes such as planet-planet scatterings are indeed responsible for
some systems with close-in giant planets.

The spin-orbit angle for multiple transiting systems is particularly interesting. 
%\textcolor{red}{
Such a multiplicity strongly suggests that the system has never
experienced chaotic processes and rather results from a
quiescent evolution history. A spin-orbit misalignment in
a multiple planetary system implies that stellar obliquities 
can vary by processes which are unconnected to planetary migrations. 
\citep[e.g., ][]{2011MNRAS.412.2790L, 2012arXiv1209.2435R}
%}
\citet{2012Natur.487..449S} measured the spin-orbit
relation for the Kepler-30 system by a precise modeling of spot-crossing
events during planetary transits and found a good spin-orbit alignment
in the system, supporting the notion that the system has evolved
quiescently in the disk.

In this letter, we focus on the KOI-94 system. This system, listed in
the earliest Kepler Object of Interest (KOI) list
\citep{2011ApJ...728..117B, 2012arXiv1202.5852B}, has four transiting
planet candidates around a relatively bright late F star 
%\textcolor{red}{
(with the Kepler magnitude of $K_p=12.205$). 
%}
The orbital period and radius $R_p$ normalized by the stellar radius $R_s$ for each
candidate are summarized in Table \ref{table0}.  Indeed KOI-94 is a
remarkable system that experienced a ``double transit'' event due to the
simultaneous transit of two planets on the stellar disk, and a
``planet-planet eclipse'' due to the partial overlap of the two planets
during the transiting phase. As we describe below, a joint analysis of
our Subaru radial velocity data and the Kepler lightcurve provides a
unique opportunity to tightly constrain the configuration of the
multi-transiting planetary system.

%The compactness of the system including a Jovian planet, KOI-94.01,
%suggests some migration history of the planets and the measurement of
%the RM effect gives us an insight into the origin of this unique system.
%We here report the first detection of the RM effect for the multiple
%transiting system, and also discuss the mutual inclination between
%KOI-94.01 and 94.03 with a lucky planet-planet crossing event, in which
%the two planets overlap on the stellar disk during the transit.

%%%%%%%%%%%%%%%%%%%
\begin{table}[t]
\caption{
Properties of Planetary Candidates in KOI-94
}\label{table0}
\begin{center}
\begin{tabular}{rcc}
\hline
Candidate & Orbital Period (days) & $R_p/R_s$\\\hline\hline
KOI-94.01& $22.343000\pm 0.000011$ & $0.06856\pm 0.00012$\\
94.02& $10.423707\pm 0.000026$ & $0.02544\pm 0.00012$\\
94.03& $54.31993\pm 0.00012$ & $0.04058\pm 0.00013$\\
94.04& $3.743245\pm 0.000031$ & $0.01045\pm 0.00019$\\
\hline
\end{tabular}
\end{center}
\end{table}
%%%%%%%%%%%%%%%%%%%

%Later sections are organized as follows. We briefly describe our observations with the 
%Subaru telescope and data reduction in Section \ref{s:sec2}. 
%Data analysis including our fitting procedures is given in Section \ref{s:sec3}. 
%Section \ref{s:sec5} will be devoted to discussion and summary. 

%%%%%%%%%%%%%%%%%%%%%%%%%%%%%%%%%%%%%%%%%%%%%%%%%%%%%%%%%%%%%%%%%%%%%%
\section{Observations and Data Reduction \label{s:sec2}}\label{s:sec2}

Due to the long orbital distance of KOI-94.01 as a transiting planet,
the transit duration is considerably long (about 6.7 hours), which makes
it very challenging to observe a complete transit within a night by a
ground-based telescope.  A fortunate opportunity was given on 2012
August 10 (UT), when a complete transit of KOI-94.01 was observed by
High Dispersion Spectrograph (HDS) installed on the Subaru 8.2 m
telescope.  We also obtained out-of-transit spectra on 2012 July 1, 2,
3, 4, August 8, 9 (UT), %September 5, 6 (UT), 
to determine the RV baseline and
derive the RV semi-amplitude of the host star.  In order to maximize the
RV precision, we adopted the I2a setup, and utilized the image slicer
\citep{2012PASJ...64...77T}
with the Iodine cell for the precise RV calibration, achieving the
spectral resolution of $R\sim 110,000$.

We reduced the raw data with the standard IRAF procedure and extracted
1-dimensional (1D) spectra.  With a 20-minutes exposure, we typically
achieved a signal-to-noise (S/N) of $80-90$ per pixel of the 1D spectra.
A high S/N template spectrum (S/N$\sim 150$) of KOI-94 was obtained
without the Iodine cell for the RV analysis.  This template spectrum was
then subject to deconvolution of the instrumental profile (IP), which
was reproduced by taking the flat lamp spectrum transmitted through the
Iodine cell.  Using this deconvolved template representing the intrinsic
stellar spectrum, we extracted the RV for each of the spectra by the RV
analysis routine developed by \citet{2002PASJ...54..873S}. The resulting
relative RV's (after corrected for the motion of the Earth), along with
their uncertainties are summarized in Table \ref{hyo1}.  The typical RV
uncertainty is $\sim $10 m s$^{-1}$.

%%%%%%%%%%%%%%%%%%%%%%%%%%%%%%%%%%%%%%%%%%%%%%%%%%%%%%%%%%%%%%%%%%%%%%
\begin{table}[tb]%[htb]
\caption{%A part of radial velocities measured with Subaru/HDS. 
Radial velocities measured with Subaru/HDS. 
An arbitrary RV offset is subtracted in the data. 
%Full version is available as an online material.
}\label{hyo1}
%\begin{center}
\begin{tabular}{lcc}
\hline
Time [BJD (TDB)]  & Relative RV [m~s$^{-1}$] & Error [m~s$^{-1}$]\\
\hline\hline
%\begin{document}$\cdots$ & $\cdots$ & $\cdots$ \\
2456109.86607 & -16.9 & 14.1\\ 
2456110.05250 & -14.0 & 16.1\\ 
2456110.84846 & -13.4 & 13.1\\ 
2456111.09425 & -24.2 & 10.7\\ 
2456111.87151 & -35.5 & 11.1\\ 
2456112.08319 & -8.2 & 10.9\\ 
2456112.85317 & -32.7 & 12.7\\ 
2456112.89785 & -31.4 & 12.1\\ 
2456147.97040 & 0.8 & 9.8\\ 
2456148.96781 & 9.7 & 13.4\\ 
2456149.75308 & -5.2 & 9.6\\ 
2456149.76802 & -13.6 & 9.5\\ 
2456149.78275 & 4.1 & 8.9\\ 
2456149.79747 & 14.2 & 8.7\\ 
2456149.81220 & 0.5 & 9.6\\ 
2456149.82692 & 15.1 & 10.1\\ 
2456149.84163 & 12.5 & 9.6\\ 
2456149.85635 & 16.2 & 9.6\\ 
2456149.87107 & 13.5 & 9.7\\ 
2456149.88579 & 9.4 & 9.6\\ 
2456149.90050 & 9.9 & 8.8\\ 
2456149.91522 & -7.1 & 9.8\\ 
2456149.92994 & -13.6 & 9.4\\ 
2456149.94466 & -16.4 & 10.0\\ 
2456149.96175 & -20.8 & 9.3\\ 
2456149.97647 & -34.6 & 8.6\\ 
2456149.99119 & -45.8 & 9.4\\ 
2456150.00591 & -44.6 & 9.2\\ 
2456150.02063 & -43.2 & 9.2\\ 
2456150.03536 & -44.0 & 9.2\\ 
2456150.05008 & -28.3 & 9.8\\ 
2456150.06480 & 9.9 & 10.4\\ 
2456150.07952 & -3.6 & 10.1\\ 
\hline
\end{tabular}
%\end{center}
\end{table}
%%%%%%%%%%%%%%%%%%%%%%%%%%%%%%%%%%%%%%%%%%%%%%%%%%%%%%%%%%%%%%%%%%%%%%

%%%%%%%%%%%%%%%%%%%%%%%%%%%%%%%%%%%%%%%%%%%%%%%%%%%%%%%%%%%%%%%%%%%%%%
\section{Analysis and Results\label{s:sec3}}\label{s:sec3}
\subsection{Stellar Parameters \label{parameters}}

Since KOI-94 is not a confirmed planetary system as of 2012 September
and no spectroscopic measurement has been published, we ourselves
estimate the basic stellar parameters 
from the template spectrum of KOI-94 used for the RV template.
Following \citet{2002PASJ...54..451T, 2005PASJ...57...27T}, we estimate
the stellar atmospheric parameters ($T_\mathrm{eff}$, the surface
gravity $\log g$, metalicity [Fe/H], and microturbulence dispersion
$\xi$), based on the equivalent widths of Fe I and Fe II lines located
at $5000-7600~\mathrm{\AA}$ wavelength region.  The result of the
measurement is shown in Table \ref{hyo2} (A).

We also extract the projected rotational velocity $V\sin I_s$ by
modeling the line profiles of KOI-94's template.  By convolving
theoretical line profiles for the same type star with the rotation plus
macroturbulence broadening kernel and the IP, we fit the line profiles
around $6100~\mathrm{\AA}$. 
%\textcolor{red}{
In doing so, we assume that the macroturbulence dispersion $\zeta$
is $4.5\pm 1.1$ km s$^{-1}$, based on the empirical relation by 
\citet{2005ApJS..159..141V}.
%}
The resulting best-fit value and its
uncertainty for $V\sin I_s$ are also shown in Table \ref{hyo2} (A).
This procedure and validity of our measurement of $V\sin I_s$ are
described by \citet{2012ApJ...756...66H} in detail.

Given the atmospheric parameters, we estimate the mass and age of the
host star.  We here adopt the Yonsei-Yale (Y$^2$) isochrone model
\citep{2001ApJS..136..417Y}.
%Based on this model and
%assuming Gaussian distributions for $T_\mathrm{eff}$ and $\log g$
%with the peaks and widths correspond to the central values and 
%uncertainties of estimated $T_\mathrm{eff}$ and $\log g$, 
Based on this model, we compute the best-fit values and their
uncertainties for the mass $M_s$, radius $R_s$, and age of KOI-94.  Table \ref{hyo2}
(A) also shows the result of those estimates.

%We note that there exists a possible stellar companion located at $\sim 0.6^{\prime\prime}$ 
%away from the main star based on the high resolution imaging by
%Subaru/HiCIAO (Takahashi et al. in preparation).
%However, the companion is faint enough compared to the main star (with the flux contrast 
%of $\sim 8\times10^{-4}$ in H-band) so that we can safely neglect the contamination 
%from the companion to the spectra.

\subsection{Modeling and Fitting of the Observed RV's\label{model}}

%%%%%%%%%%%%%%%%%%%%%%%%%%%%%%%%%%%%%%%%%%%%%%%%%%%%%%%%%%%%%%%%%%%%%%
\begin{figure}[ptb]
 \begin{center}
 %\FigureFile(85mm,85mm){koi94_fit_orbit.eps} %160mm,160mm or 85mm,85mm
 %\FigureFile(85mm,85mm){koi94_fit_RM.eps} %160mm,160mm or 85mm,85mm
  \includegraphics[width=8.5cm,clip]{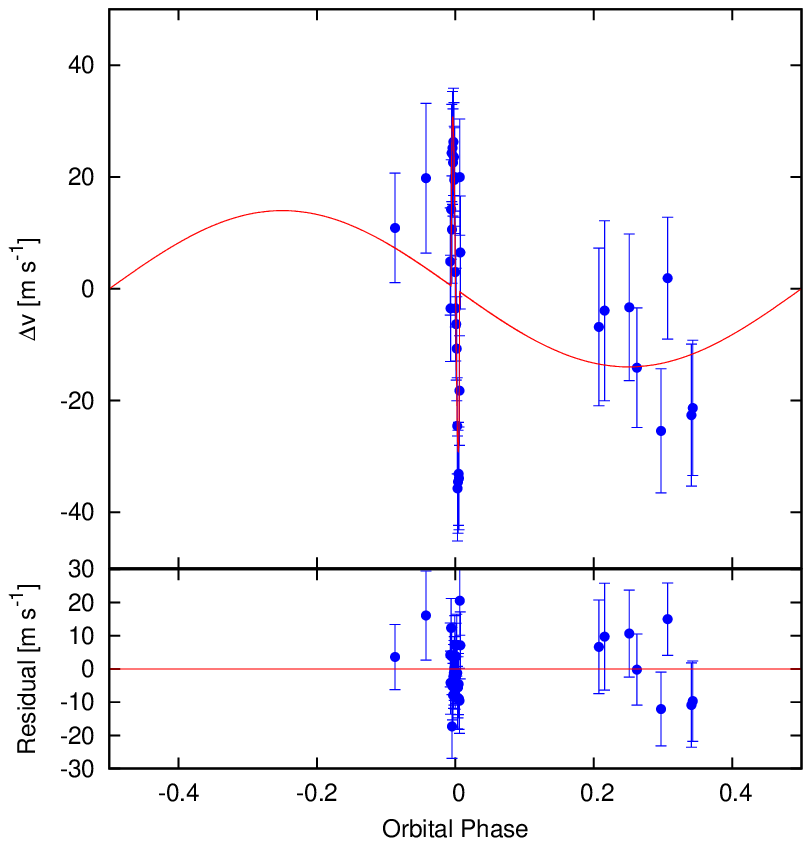}
  \includegraphics[width=8.5cm,clip]{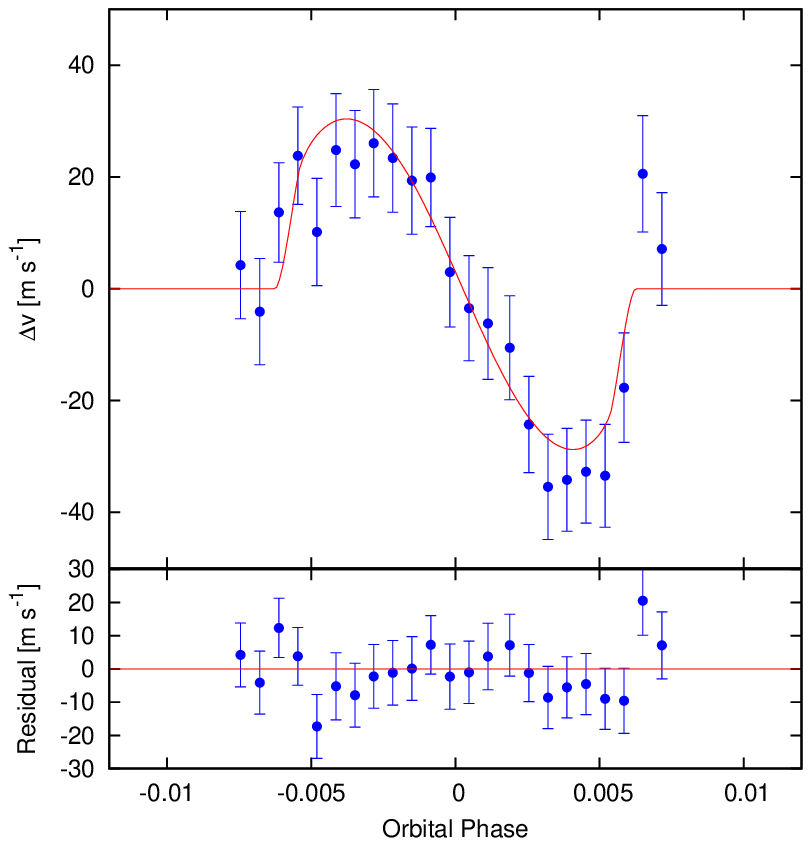}
 \end{center}
  \caption{
  (Upper) 
  Phase-folded RV data of KOI-94, obtained with Subaru/HDS. 
  The best-fit model curve is shown in red.
  (Lower) 
  The RV variation around the planetary transit of KOI-94.01 (blue)
  and its best-fit model (red solid line). 
  The Keplerian orbit is subtracted from the RV's. 
  }\label{fig1}
\end{figure}
%%%%%%%%%%%%%%%%%%%%%%%%%%%%%%%%%%%%%%%%%%%%%%%%%%%%%%%%%%%%%%%%%%%%%%

%%%%%%%%%%%%%%%%%%%%%%%%%%%%%%%%%%%%%%%%%%%%%%%%%%%%%%%%%%%%%%%%%%%%%%
%\begin{table}[tb]
%\caption{Best-fit system parameters.}\label{hyo3}
%\begin{center}
%\begin{tabular}{lc}
%\hline
%Parameter& best-fit value \\\hline\hline
%$K$  [m s$^{-1}$]& $14.2_{-4.9}^{+5.0}$  \\%&  1494.0 $\pm$ 9.5 \\
%$V\sin I_s$ [km s$^{-1}$]& $8.00\pm 0.75$ \\%& $18.4\pm0.8$ &\\
%$\la$  [$^\circ$]&$-7_{-11}^{+12}$ \\%& $37.4\pm 2.2$ &\\
%$\tilde{\chi}^2$ &0.84 \\%&1.00 (fixed) &\\
%\hline
%\end{tabular}
%\end{center}
%\end{table}
%%%%%%%%%%%%%%%%%%%%%%%%%%%%%%%%%%%%%%%%%%%%%%%%%%%%%%%%%%%%%%%%%%%%%%

The upper panel of Figure \ref{fig1} plots the RV data phase-folded with
the orbital period of KOI-94.01 (i.e., $\sim 22.34$ days), and the lower
panel is the zoomed-in version around the transit phase after
subtracting the orbit of KOI-94.01.
%With a glimpse at the RV variation during the transit of KOI-94.01, 
The RV variation during the transit of KOI-94.01 clearly indicates that
the RM effect is securely detected, and that the planetary orbit 
is prograde with respect to the host star's equator.

Following \citet{2011PASJ...63L..57H} and \citet{2011PASJ...63L..67N},
we fit the observed RV's assuming that the RV's can be approximated by
the RV variation due to Keplerian motion of KOI-94.01 and the velocity
anomaly due to the RM effect. Because of the lack of RV data points, we
here neglect the impacts of KOI-94.02, 94.03, 94.04 on RV variations.
Based on the transit depths of planetary candates reported by Kepler and
using a simple relation between the planet mass and radius, $M_p /
M_\oplus = (R_p / R_\oplus)^{2.06}$, where $M_\oplus$ and $R_\oplus$ are
respectively the Earth mass and radius \citep{2011ApJS..197....8L}, we
roughly estimate the planet mass and compute the RV semi-amplitude
exerted by each candidate; the ratios of RV semi-amplitudes for
KOI-94.02, 03, and 04 to the semi-amplitude for KOI-94.01 are estimated
to be 0.17, 0.25, and 0.04, respectively.  Considering that the maximum
RV variation in Figure \ref{fig1} is approximately $30$ m s$^{-1}$ and
our error for each RV is no less than $10$ m s$^{-1}$, one-planet
approximation for the RV variation of KOI-94 is not a bad approximation.
Precise measurements of the impacts KOI-94.02, 94.03, and 94.04 are
beyond the scope of this letter and will be discussed together with
further follow-up observations.

The RV variation including a transit phase is usually modeled as
%%%%%%%%%%%%%%%%%%%%%%
\begin{eqnarray}
\label{RVmodel}
%V_\mathrm{model}=K[\cos(f+\varpi)+e\cos(\varpi)]+\D v_\mathrm{RM}+\gamma_\mathrm{offset},
V_\mathrm{model}=K[\cos(f+\varpi)+e\cos(\varpi)]+\D v_\mathrm{RM},
\end{eqnarray}
%%%%%%%%%%%%%%%%%%%%%%
where $K$, $e$, and $\varpi$ are the RV semi-amplitude, orbital
eccentricity, and argument of periastron, respectively.  Again, because
of the lack of RV data points and the fact that KOI-94 is a densely
packed multiple planetary system with two giant planets being within
$\sim 0.3$ AU from the host star, we decided to fix the orbital
eccentricity as zero, and put a rough constraint on the RV
semi-amplitude for KOI-94.01.  Future long-term monitoring will %clarify
the possibility of non-zero eccentricity.  improve the analysis with
non-vanishing eccentricity.

%%%%%%%%%%%%%%%%%%%%%%%%%%%%%%%%%%%%%%%%%%%%%%%%%%%%%%%%%%%%%%%%%%%%%%
\begin{table}[tb]
\caption{System Parameters}\label{hyo2}
\begin{center}
%\begin{tabular}{lcl}
\begin{tabular}{lc}
\hline
Parameter& Value\\\hline\hline
(A) {\it Spectroscopic Parameters}&\\
$T_\mathrm{eff}$  (K)& 6116 $\pm$ 30 \\ %& \citet{2005PASJ...57...27T}\\
$\log g$ (dex) & $4.123\pm0.055$ \\ %& \citet{2005PASJ...57...27T}\\
$[\mathrm{Fe/H}]$ (dex) & $-0.01\pm 0.04$ \\ %& \citet{2005PASJ...57...27T}\\
$\xi$ (km s$^{-1}$) & $1.29\pm 0.20$ \\ %& \citet{2005PASJ...57...27T}\\
$V\sin I_s$ (km s$^{-1}$) & $7.33\pm 0.32$ \\ %& \\
$M_s$ ($M_\odot$) & $1.25_{-0.04}^{+0.03}$ \\ % & Y$^2$ isochrone\\
$R_s$ ($R_\odot$) & $1.61_{-0.12}^{+0.11}$\\
age (Gyr) & $3.9_{-0.2}^{+0.3}$ \\ %& Y$^2$ isochrone\\
\hline
(B) {\it Orbital and RM Parameters}&\\
$K$  (m s$^{-1}$)& $14.3\pm 4.9$ \\
$V\sin I_s$ (km s$^{-1}$)& $8.01_{-0.73}^{+0.72}$\\
$\la$  ($^\circ$)& $-6_{-11}^{+13}$ \\
$\gamma_\mathrm{offset}$ (m s$^{-1}$) & $-9.9_{-2.4}^{+2.5}$ \\
$\tilde{\chi}^2$ &0.80 \\
\hline
(C) {\it Double Transit Parameters}&\\
$T_c^{(1)}$ (BJD) & $2455211.51363 \pm 0.00024$\\
$T_c^{(3)}$ (BJD) & $2455211.51790 \pm 0.00062$\\
$u_1$ & $0.10 \pm 0.06$ \\
$u_2$ & $0.61 \pm 0.08$ \\
$\delta$ ($^\circ$) & $-1.15 \pm 0.55$\\
\hline
\end{tabular}
\end{center}
\end{table}
%%%%%%%%%%%%%%%%%%%%%%%%%%%%%%%%%%%%%%%%%%%%%%%%%%%%%%%%%%%%%%%%%%%%%%

In order to model the RM velocity anomaly $\D v_\mathrm{RM}$, we employ
the analytic formula by \citet{2011ApJ...742...69H}. The analytic
formula is derived for the case that RV's are extracted by fitting the
stellar template spectrum taken outside of the transit to the distorted
spectrum during a transit, and therefore is more suitable for our
calibration of the RM velocity anomaly than the classical formula
derived by taking the intensity-weighted center of the distorted line
profile \citep[e.g., ][]{2005ApJ...622.1118O}. For the stellar
parameters required for the RM formula, we adopt the macroturbulence
dispersion of $\zeta=4.5$ km s$^{-1}$, and intrinsic Gaussian and
Lorentzian line widths of $\beta=2.4$ km s$^{-1}$ and $\gamma=1.0$ km
s$^{-1}$, respectively, based on the empirical estimates by
\citet{2011ApJ...742...69H}.  As for the limb-darkening, we employ the
quadratic limb-darkening law with $u_1=0.40$ and $u_2=0.31$ based on the
table by \citet{2004A&A...428.1001C}.  Also, we here adopt
$R_p/R_s=0.06856$, $a/R_s=26.1$, and $I_o=89.33^\circ$ for the transit
parameters of KOI-94.01, which are the values provided by the Kepler
team.

We fit the observed RV's using Markov Chain Monte Carlo (MCMC)
algorithm.  The $\chi^2$ statistics in our case is
%%%%%%%%%%%%%%%%%%%%%%
\begin{eqnarray}
\chi^2 = \sum_{i} \frac{(V_{\mathrm{model}, i} - V_{\mathrm{obs}, i})^2}{\sigma_i^2},
\end{eqnarray}
%%%%%%%%%%%%%%%%%%%%%%
where $V_{\mathrm{model}, i}$ and $V_{\mathrm{obs}, i}$ are the $i-$th
modeled and observed RV's, and $\sigma_i$ is its error.
%We run 1,000,000 chains and compute the best-fit value as the median of the
%chain and take the lower and upper $34\%$ as $1\sigma$ errors. 
%We try several different cases for fitting the RV data. 
The remaining fitting parameters are $K$, the spin-orbit angle
$\lambda$, $V\sin I_s$, and the RV offset (zero-point)
$\gamma_\mathrm{offset}$ of our dataset listed in Table \ref{hyo1}.  
The resulting best-fit values and $1\sigma$ errors are summarized in Table \ref{hyo2} (B).  
The red curves in Figure \ref{fig1} indicate the best-fit models for the whole
orbit (upper) and the RM effect (lower), respectively.

The best-fit value of $V\sin I_s$ purely estimated by the RM velocity
amplitude ($V\sin I_s = 8.01_{-0.73}^{+0.72}$ km s$^{-1}$) agrees well with the
spectroscopically measured value ($V\sin I_s = 7.33\pm 0.32$ km
s$^{-1}$), validating our measurement and modeling of the RM effect.
One concern in our analysis is that since we neglected the impacts of
the other three planetary candidates, we might have obtained a
systematically biased estimate for the RV semi-amplitude $K$ of
KOI-94.01.  A spurious RV baseline during the transit phase leads to a
systematically biased estimate for the spin-orbit angle $\la$, and
sometimes makes an aligned system look misaligned, or vice versa.  Thus,
%\textcolor{red}{
we perform the following test; we remove the RV data on the transit night,
and fit the remaining RV's with $K$ and $\gamma_\mathrm{offset}$ being
free. Then, using the best-fit value of $K$, we fit the RV's during the transit
night in order to derive $\lambda$ and $V\sin I_s$ for the fixed $K$. 
This treatment can remove possible systematics 
caused by instrumental effects, the impact of other planets, and/or 
non-zero eccentricity. 
As a result, we find $V\sin I_s=8.02_{-0.74}^{+0.86}$ km s$^{-1}$ and $\la=-10^{+13}_{-11}$ deg,
implying a good spin-orbit alignment again. 
%}
%we intentionally change and fix the $K$ value at (a) $10$ m~s$^{-1}$ and
%(b) $25$ m~s$^{-1}$ and see if the spin-orbit angle $\la$ varies by
%refitting the observed RV's.  Indeed, we obtain (a)
%$\la=-10^{+11}_{-10}$ deg and (b) $\la=0^{+13}_{-11}$ deg, respectively,
%implying a good spin-orbit alignment for both cases.  
Based on this test, 
we conclude that the measurement of the spin-orbit angle $\la$ is
robust and the orbital axis of KOI-94.01 is almost parallel to the
projected stellar spin axis.

Substituting the best-fit value of $K$ and our estimate for the stellar
mass $M_s$, we obtain $\sim 0.23M_J$ for the planet mass of KOI-94.01.  
%\textcolor{red}{
This mass is slightly small compared to the theoretical expectation 
for a planet having a radius of $\sim 9.25R_\oplus$ \citep{2012A&A...541A..97M},
but is rather similar to low-density planets like Kepler-9c \citep{2010Sci...330...51H}. 
%}
In our analysis, however, we completely neglected the
impacts of the other three planet candidates so that this estimate of
the planet mass more or less has systematics.  Again, a long-term RV
monitoring is required in order to put tight constraints on masses and
eccentricities of all the planet candidates in this system.

%\subsection{Mutual Inclination between Planets \label{mutual}}

%%%%%%%%%%%%%%%%%%%%%%%%%%%%%%%%%%%%%%%%%%%%%%%%%%%%%%%%%%%%%%%%%%%%%%
\section{Discussion and Summary\label{s:sec5}}\label{s:sec5}

%%%%%%%%%%%%%%%%%%%%%%%%%%%%%%%%%%%%%%%%%%%%%%%%%%%%%%%%%%%%%%%%%%%%%%
\begin{figure}[ptb]
 \begin{center}
  %\FigureFile(85mm,85mm){koi94_fit_doubletransit.eps} %160mm,160mm or 85mm,85mm
  \includegraphics[width=8.5cm,clip]{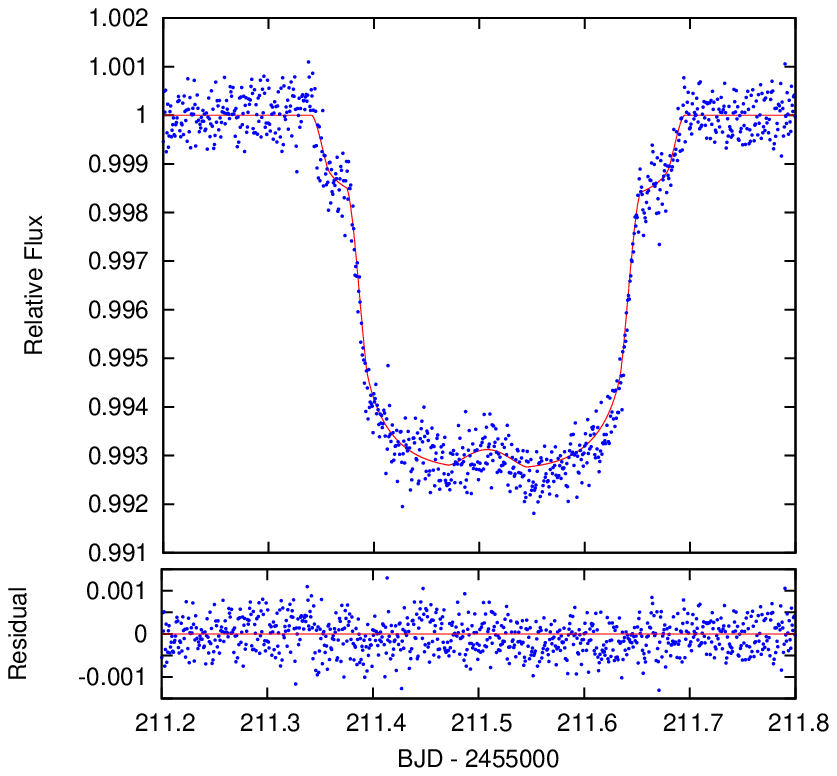} 
  \includegraphics[width=8.5cm,clip]{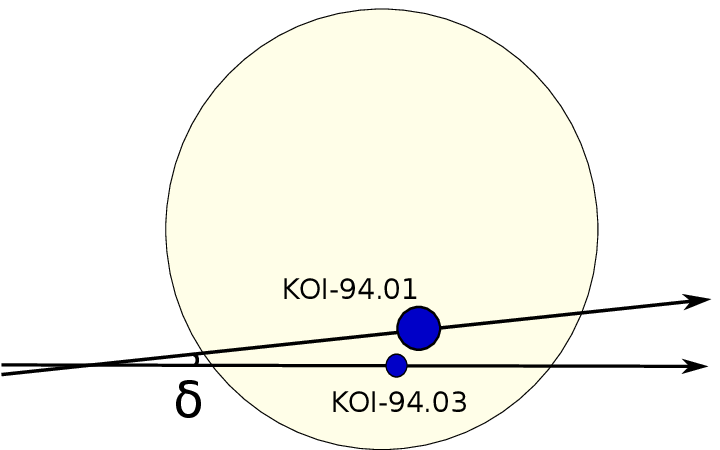} 
 \end{center}
  \caption{ 
  (Upper)
  The public lightcurve taken by Kepler around
  $\mathrm{BJD}=2455211.5$.  In addition to the simultaneous double
  transit event, there is a bump around the transit center of KOI-94.01,
  which most likely represents the planet-planet eclipse.
  (Lower)
  The definition of the mutual inclination $\delta$ between KOI-94.01 and 94.03. 
  The orbits of the two planets are projected onto the sky plane. This is just a schematic
  description and does not reflect the the real planet sizes and angle between the planetary orbits. 
  }\label{fig3}
\end{figure}
%%%%%%%%%%%%%%%%%%%%%%%%%%%%%%%%%%%%%%%%%%%%%%%%%%%%%%%%%%%%%%%%%%%%%%

While we have seen that KOI-94.01's orbital axis is aligned with the
stellar rotation axis, it is not evident that the orbits of the other
planet candidates are also neatly aligned with the stellar spin.
Fortunately, however, we found a lucky lightcurve in which a ``double
transit'' event is observed in Kepler's public lightcurve. 
The upper panel of Figure
\ref{fig3} shows the normalized lightcurve of KOI-94 around
$\mathrm{BJD}=2455211.5$. In this figure, the first transit event starts
around $\mathrm{BJD}=2455211.35$, and after a while, it shows the
second, deeper dimming. According to the public Kepler planet candidate
ephemeris, the first shallow dimming represents the transit of KOI-94.03
and the second deeper one corresponds to a transit of KOI-94.01.
Interestingly enough, there is a bump around the transit center with
height of $\sim 0.0003$ in terms of the relative flux.  Similar events
have been frequently seen when the transiting planet crosses a starspot
\citep{2011ApJ...733..127S}. 
%\textcolor{red}{
We checked the out-of-transit lightcurve (long-cadence data) and computed the periodogram
in order to look for a signal induced by starspots. Consequently, we found
a weak peak around the period of $\sim 11$ days. However, the flux variability 
in the lightcurve \citep[as is defined in][]{2012ApJ...756...66H} is about $0.02\%$
so that the size of the bump seen in Figure \ref{fig3} cannot be explained by 
a spot-crossing event. Considering the timing and size of the bump, 
the most likely scenario would be the eclipse of the planet by another transiting planet 
\citep[planet-planet eclipse, also called ``mutual event", ][]{2010arXiv1006.3727R}.
%} 
Since KOI-94.03 has the
longer orbital distance from the host star, a part of the inner planet,
KOI-94.01, is occulted by the outer transiting planet, KOI-94.03, during
the eclipse event represented by the bump.

Since the impact parameters of the two transiting planets are
fixed from the transit photometries, 
the only parameter to determine the relative planetary orbits on the stellar disk
and describe the bump in Figure \ref{fig3} is the
mutual inclination between the two transiting planets. If the mutual
inclination of the two planets is considerably large, we expect no
planet-planet eclipse. The timing and shape of the anomalous bump in
Figure \ref{fig3} put a strong constraint on the mutual inclination.

%%%%%%%%%%%%%%%%%%%%%%%%%%%%%%%%%%%%%%%%%%%%%%%%%%%%%%%%%%%%%%%%%%%%%%
%\begin{figure}[ptb]
 %\begin{center}
  %\FigureFile(85mm,85mm){koi94_fit_doubletransit.eps} %160mm,160mm or 85mm,85mm
  %\includegraphics[width=8.5cm,clip]{double_transit_skematic.eps} 
 %\end{center}
  %\caption{
  %The definition of the mutual inclination $\delta$ between KOI-94.01 and 94.03. 
  %The orbits of the two planets are projected onto the sky plane. This is just a schematic
  %description and does not reflect the the real planet sizes and angle between the planetary orbits. 
  %}\label{fig4}
%\end{figure}
%%%%%%%%%%%%%%%%%%%%%%%%%%%%%%%%%%%%%%%%%%%%%%%%%%%%%%%%%%%%%%%%%%%%%%

%%%%%%%%%%%%%%%%%%%%%%%%%%%%%%%%%%%%%%%%%%%%%%%%%%%%%%%%%%%%%%%%%%%%%%
%\begin{table}[tb]
%\caption{Best-fit Parameters for the Double Transit Event.}\label{hyo5}
%\begin{center}
%\begin{tabular}{lc}
%\hline
%Parameter& best-fit value \\\hline\hline
%$T_c^{(1)}$ & $2455211.51363 \pm 0.00024$\\
%$T_c^{(3)}$ & $2455211.51790 \pm 0.00062$\\
%$u_1$ & $0.10 \pm 0.06$ \\
%$u_2$ & $0.61 \pm 0.08$ \\
%$\delta$ ($^\circ$) & $-1.15 \pm 0.55$\\
%\hline
%\end{tabular}
%\end{center}
%\end{table}
%%%%%%%%%%%%%%%%%%%%%%%%%%%%%%%%%%%%%%%%%%%%%%%%%%%%%%%%%%%%%%%%%%%%%%
We define the mutual inclination projected onto the sky as $\delta$ (see
the lower panel of Figure \ref{fig3}) and model the anomalous bump by a simple 
geometric calculation as a function of time and $\delta$. 
%\textcolor{red}{
The mutual inclination $\delta$ sensitively changes the duration of
the planet-planet eclipse, and can be constrained by the shape of the
bump around the bottom of the transit lightcurve; the resulting bump
is longer for nearly prograde ($\delta \approx 0$), and shorter for
retrograde planets ($\delta \approx \pi$).
%}
By modeling the double
transit event as a sum of two single transit lightcurves and the bump
function (which is expressed analytically as an area fraction of the
overlapping region by the two transiting planets), we fit the observed
lightcurve by MCMC algorithm. The free parameters in our fit are the
mid-transit times $T_c^{(1)}$ and $T_c^{(3)}$ for KOI-94.01 and
KOI-94.03, the limb-darkening parameters $u_1$ and $u_2$ for the
quadratic limb-darkening law\footnote{These coefficients should be
different from the ones used for the modeling of the RM effect since
these represent the limb-darkening coefficients for the Kepler band.},
and the projected mutual inclination $\delta$.  The other transit
parameters (i.e., $R_p/R_s$, $a/R_s$, and $i_o$) for both of 94.01 and
94.03 are fixed at the values delivered by the Kepler team assuming
circular orbits for both of the candidates. Table \ref{hyo2} (C) shows
the best-fit values and their uncertainties of our fitting.  The red
solid line in the upper panel of Figure \ref{fig3} indicates the best-fit
model and the residual of observed data from the best-fit model is shown
at the bottom.  
The orbital planes of the two transiting planets are
remarkably well aligned, which means that host star's spin axis and planetary
axes of both KOI-94.01 and 94.03 are all well aligned.
%In order to discuss the spin-orbit relations for
%KOI-94.02 and 94.04 as well, we have checked public transit lightcurves for
%those candidates, but we could not locate any planet-planet crossing event
%with the currently available data. Future announcement of the new data 
%would enable us to discuss the whole spin-orbit configuration of the system. 

Like the Kepler-30 system, the spin-orbit alignment in the KOI-94 system
implies that those planets in this multiple transiting system have
experienced a quiescent migration process rather than having been
carried to the present locations by dynamical scattering processes or
long term perturbations by outer objects.  This has also ruled out the
possibility that an earlier magnetic interaction between the star and
disk causes a gradual misalignment of the stellar spin from the disk
plane \citep{2011MNRAS.412.2790L} at least for this system.  On the
other hand, our result should reinforce the hypothesis that spin-orbit
misalignments are only seen for ``isolated'' hot Jupiters, which may
have migrated by chaotic processes \citep{2012Natur.487..449S}. 
%, but further observations of both single and multiple transiting systems are required to
%reach a unified picture of the evolution history of planetary systems. 

%\textcolor{red}{
Finally, we note that KOI-94 has the effective 
temperature of $T_\mathrm{eff}=6116 \pm 30$ K, which 
suggests that the star most likely has a convective envelope rather than
being fully covered by a thick radiative envelope as seen for massive
(hotter) stars \citep{2001ApJ...556L..59P}. Thus, this system does not provide us the chance to test
the recent theoretical work by \citet{2012arXiv1209.2435R}, stating that the stellar spins of massive stars 
with radiative envelopes and convective cores may spontaneously change directions
due to the internal gravity waves (IGW) generated around the border
between the radiative envelope and convective core,
regardless the presence of planets.  
A future observation of the RM effect for a multiple transiting system
with a massive host star will confirm or refute this intriguing scenario. 
%}

%%%%%%%%%%%%%%%%%%%%%%%%%%%%%%%%%%%%%%%%%%%%%%%%%%%%%%%%%%%%%%%%%%%%%%
\acknowledgments 

This letter is based on data collected at Subaru Telescope, which is
operated by the National Astronomical Observatory of Japan.  We are very
grateful to the support for our Subaru HDS observations by Akito
Tajitsu, a support scientist for the Subaru HDS.  We express special
thanks to Hiroki Harakawa and Eric Gaidos for exchanging the observation time 
for obtaining additional RV data, and to Roberto Sanchis-Ojeda and Atsushi Taruya 
for a helpful discussion
and important suggestions on this manuscript. The data analysis was in
part carried out on common use data analysis computer system at the
Astronomy Data Center, ADC, of the National Astronomical Observatory of
Japan.  T.H.\ is supported by Japan Society for Promotion of Science
(JSPS) Fellowship for Research (DC1: 22-5935).  T.H. and
N.N. acknowledge a support by NINS Program for Cross-Disciplinary Study.
N.N. is supported by Grant-in-Aid for Research Activity Start-up
No. 23840046.  M.T.\ is supported by the Ministry of Education, Science,
Sports and Culture, Grant-in-Aid for Specially Promoted Research,
22000005.  Y.S. gratefully acknowledges support from the Global
Collaborative Research Fund ``A World-wide Investigation of Other
Worlds'' grant and the Global Scholars Program of Princeton University,
the Grant-in-Aid No. 24340035 by JSPS.  We acknowledge the very
significant cultural role and reverence that the summit of Mauna Kea has
always had within the indigenous people in Hawai'i.
Finally, we deeply acknowledge the referee, Joshua Winn,
for his important suggestions to improve the manuscript.

%%%%%%%%%%%%%%%%%%%%%%%%%%%%%%%%%%%%%%%%%%%%%%%%%%%%%%%%%%%%%%%%%%%%%%
%\appendix
%\section{Bump Function \label{ap:bump}}\label{ap:bump}
%Here we present the analytic form of the bump function.  In the derivation, we assume that the orbital eccentricities of 
%KOI-94.01 and KOI-94.03 are zero and the planetary trajectories on the stellar disk are approximated as lines. In this case, 

%%%%%%%%%%%%%%%%%%%%%%%%%%%%%%%%%%%%%%%%%%%%%%%%%%%%%%%%%%%%%%%%%%%%%%
%\bibliography{hirano2012_reference}
%\end{document}

%%%%%%%%%%%%%%%%%%%%%%%%%%%%%%%%%%%%%%%%%%%%%%%%%%%%%%%%%%%%%%%%%%%%%%

%%%%%%%%%%%%%%%%%%%%%%%%%%%%%%%%%%%%%%%%%%%%%%%%%%%%%%%%%%%%%%%%%%%%%%

\end{document}